\documentclass[11pt,a4paper]{article}
\pdfoutput=1
\usepackage{jheppub} 
\usepackage{graphics,graphicx}
\usepackage{amsfonts,amsmath,amsthm,amssymb}

\title{Relationship between gravity and gauge scattering in the high energy limit}

\author[a]{Ryo Saotome}
\author[a]{Ratindranath Akhoury}
\affiliation[a]{Michigan Center for Theoretical Physics, Randall Laboratory of Physics, University of Michigan, Ann Arbor, MI 48109-1120, USA}
\emailAdd{rsaotome@umich.edu}
\emailAdd{akhoury@umich.edu}

\abstract{Investigations of high-energy graviton-graviton and gluon-gluon scattering are performed in the leading eikonal approximation for the kinematic regime of large center of mass energy and low momentum transfer. We find a double copy relation between the amplitudes of the two theories to all loop orders when, on the gauge theory side, we retain only the set of diagrams at each loop order  for which the collinear divergences cancel amongst themselves. For this to happen the color structure of all diagrams in a set must be arranged to be identical. Using standard field theoretic methods, it is shown that this relation is reflected in a similar double copy relationship between the classical shockwaves of the two theories as well.}

\begin{document}

\maketitle

\section{Introduction}

There has been a renewed interest in perturbative quantum gravity due to recent developments illuminating a relationship between gravity and gauge theory amplitudes \cite{BCJ1, BCJ2}. This relationship is thought to endure to all loop orders and would imply a much deeper relationship between gravity and gauge theories than is apparent at the Lagrangian level. It is this connection we wish to explore further in this paper in the context of a simplified approximation scheme, the eikonal approximation, relevant for high energy large $s$ and small $t$ scattering. It has been known for some time \cite{thooft} that the there is a relationship between the amplitudes in the eikonal scheme to the scattering of a quantum particle in a classical shockwave background. In an attempt to extend the double copy relation beyond the domain of scattering amplitudes, we also explore the relationship between the relevant shockwave solutions in gravity and gauge theories. Our results support the double copy conjecture on both fronts.

High-energy graviton-graviton scattering in the eikonal approximation has attracted interest because of the potential relevance of the large $s$ small $t$ regime to black hole production \cite{thooft, Veneziano1, Veneziano2, Veneziano3, Veneziano4, Veneziano5, GGM,Giddings}. Furthermore, this approximation scheme has been used to investigate gravitational infrared divergences, which have been shown to have an elegant and simple structure \cite{Weinberg, Us, Beneke}. For the purposes of this paper, a key result of these investigations, which holds beyond the eikonal approximation, is that for scattering amplitudes in perturbative quantum gravity, the collinear divergences completely cancel to all loop orders. This will be used to identify the class of gauge theory diagrams that participate in the double copy relation with the gravity amplitudes in the eikonal approximation.

In light of these new developments, as mentioned above, we have investigated whether or not it is possible to elucidate a double-copy relationship between gravity and gauge theory amplitudes in the eikonal approximation. Some progress along these lines has been made in \cite{White}, where the insertion of infrared gravitons and gluons into a tree level diagram with hard momentum exchange was investigated. The results of this investigation upheld the double copy conjecture in the infrared approximation to all loop orders. However, in this paper we will be considering high energy scattering at large $s$ and small $t$ where there is no hard momentum transfer at all.  This  situation is drastically different for a number of reasons. First, the large $s$ and $|u|$ limit is such that the number of diagrams contributing at leading order in $1/s$ is reduced greatly. Secondly, in this approximation the numerator factors do not contain any factors of the loop momenta, as these are suppressed compared to the external momenta. Thus the numerator factors are uniquely fixed in terms of the external momenta  in both the gauge theory and the gravity side. In a certain sense this makes the analysis easier. On the other hand, the arguments of \cite{BCJ1, BCJ2} do not directly apply to such a situation. Fortunately, the fact that collinear divergences in gravity cancel allows us to identify which gauge theory diagrams participate in the double copy relation.

Another aspect of the relationship between eikonalized gravity and gauge theory amplitudes that we have explored in this paper concerns classical shockwave solutions \cite{AS, DT, Jackiw}. We detail a new method to directly derive both gravity and gauge shockwaves using eikonalized field-theoretic methods, and show that due to the direct nature of this derivation these shockwaves reflect the double copy relation as well. This double copy relation is only clearly seen in a certain gauge, which seems to be a generic property of such relations. We would like to emphasize that this is one of the few instances where a double copy relation has been studied for classical solutions to the two theories in question and that in order to make this analysis we necessarily had to restrict ourselves to the eikonal regime where there is no hard momentum transfer.

A pedagogical review of eikonal graviton-graviton scattering is given in section \ref{sec:grav}. Analogous gluon-gluon scattering results and the restrictions necessary to get a double copy relation are described in section \ref{sec:qcd}. The resultant eikonal double copy relation is described in section \ref{sec:double}. In sections \ref{sec:gravshock} and \ref{sec:gaugeshock} we derive respectively the gravity and QCD shockwaves before describing their relationship in section \ref{sec:doubleshock}. In the final section we review our results.

\section{Eikonal Scattering}

In this section we present results on high-energy graviton-graviton and gluon-gluon scattering in the eikonal limit. As mentioned previously, we will work in the kinematic limit  of large $s$ and small $t$ and keep only the leading contributions. Thus, up to corrections of order $1/s$, only the $t$ channel exchange diagrams contribute. Also in keeping with the eikonal approximation, we will neglect all the loop momenta in the numerator factors, and replace the denominators of all propagators by the rule:
\begin{align}
(P + K)^2 - m^2 + i\epsilon \rightarrow 2P\cdot K +i\epsilon,
\end{align}
where, $P$ is an external momentum and $K$ denotes any combination of internal loop momenta. We will begin with a review of  graviton-graviton scattering before moving on to the case of gluon-gluon scattering.

\subsection{Gravity}
\label{sec:grav}
Let use start with a review of the gravity case in order to establish our conventions and because we will be using these results in later sections. We will be working in the deDonder gauge and the frame where $\Delta=p'-p$ is such that $\Delta^0=\Delta^z=0$, in coordinates where $\bf{p}$ is in the positive z-direction and $\bf{q}$ is in the negative z-direction. This is the center of momentum frame in the kinematic regime where we can ignore the transferred momentum compared to the incoming and outgoing momenta.
\subsubsection{Tree Level}

\begin{figure}
\begin{center}
\includegraphics[width=2.5in]{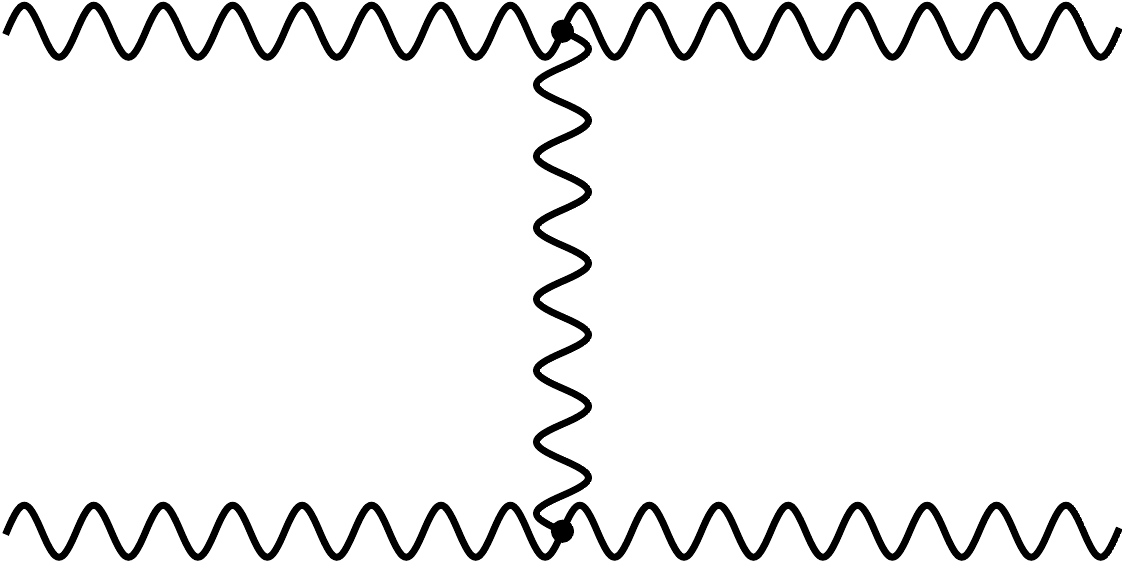}
\caption{A scattering process with a single graviton/gluon exchanged.}
\label{fig1}
\end{center}
\end{figure}

Let us first work with the simplest case where only a single graviton is exchanged (figure \ref{fig1}). In this case, the matrix element corresponding to this diagram is:
\begin{align}
\mathcal{M}_1&=(-\frac{i\kappa}{2})^2\frac{i}{2}\frac{L_{\mu\nu\alpha\beta}}{\Delta^2+i\epsilon}
\tau^{\alpha\beta\gamma\delta\sigma\eta}(p)\tau^{\mu\nu\omega\epsilon\rho\lambda}(q)
\epsilon_{\gamma\delta}(p)\epsilon_{\rho\lambda}(q)\epsilon^*_{\sigma\eta}(p)\epsilon^*_{\omega\epsilon}(q)
\end{align}
where:
\begin{align}
L_{\mu\nu\alpha\beta}=\eta_{\mu\alpha}\eta_{\nu\beta}+\eta_{\mu\beta}\eta_{\nu\alpha}-\eta_{\mu\nu}\eta_{\alpha\beta}
\end{align}
is the numerator of the de Donder gauge graviton propagator and
\begin{align}
\tau^{\mu\nu\omega\epsilon\rho\lambda}(q)\approx q^\mu q^\nu L^{\omega\epsilon\rho\lambda}
\end{align}
is the three graviton vertex in the eikonal limit. Thus, we have for the one graviton exchange amplitude:
\begin{align}
\mathcal{M}_1&=(-\frac{i\kappa}{2})^2\frac{is^2}{\Delta^2+i\epsilon}.
\label{gravTree}
\end{align}
\subsubsection{One-Loop}

\begin{figure}
\begin{center}
\includegraphics[width=2.5in]{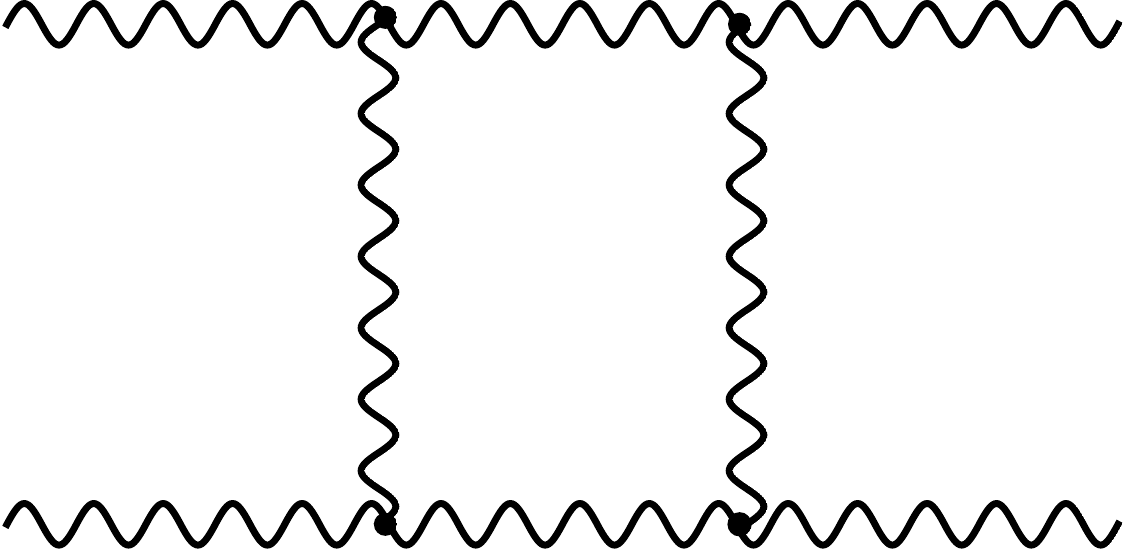}
\caption{The box diagram.}
\label{fig2}
\end{center}
\end{figure}

\begin{figure}
\begin{center}
\includegraphics[width=2.5in]{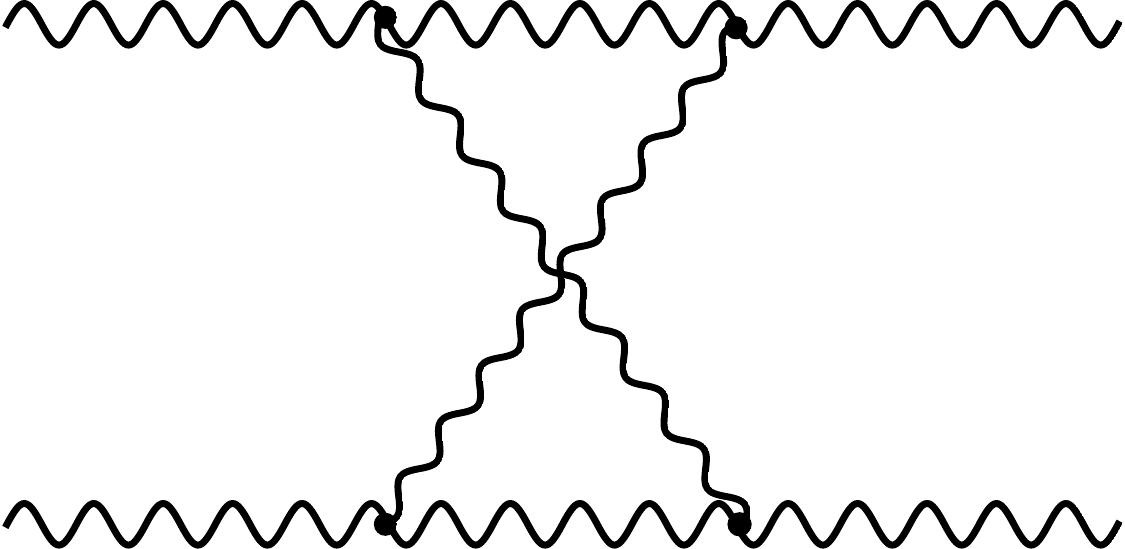}
\caption{The crossed box diagram.}
\label{fig3}
\end{center}
\end{figure}

Let us now consider the scattering process where two gravitons are exchanged. At this order there are two diagrams, the box and the crossed box (figures \ref{fig2} and \ref{fig3}). In this case, the matrix element becomes:
\begin{align}
\mathcal{M}_2&=(-\frac{i\kappa}{2})^4\frac{1}{16}L_{\gamma'\delta'\sigma'\eta'}L_{\rho'\lambda'\omega'\epsilon'}\tau^{\alpha_2\beta_2\sigma'\eta'\sigma\eta}(p)
\tau^{\alpha_1\beta_1\gamma\delta\gamma'\delta'}(p) \tau^{\mu_1\nu_1\rho'\lambda'\rho\lambda}(q)\tau^{\mu_2\nu_2\omega\epsilon\omega'\epsilon'}(q)
\nonumber \\
&\times \int \frac{d^4 k_1 d^4k_2}{(2\pi)^4}\delta(k_1+k_2+\Delta)
\frac{1}{k_1^2+i\epsilon}\frac{1}{k_2^2+i\epsilon} \frac{1}{-2p\cdot k_1 +i\epsilon}
\nonumber \\
&\times[\frac{L_{\alpha_1\beta_1\mu_1\nu_1}L_{\alpha_2\beta_2\mu_2\nu_2}}{2q\cdot k_1 +i\epsilon}+\frac{L_{\alpha_1\beta_1\mu_2\nu_2}L_{\alpha_2\beta_2\mu_1\nu_1}}{2q\cdot k_2 +i\epsilon}]
\epsilon_{\gamma\delta}(p)\epsilon_{\rho\lambda}(q)\epsilon^*_{\sigma\eta}(p)\epsilon^*_{\omega\epsilon}(q)
\end{align}
where the sum on the third line is the sum over the two diagrams. It is apparent that even at the one-loop level there are a huge amount of internal space-time indices. The situation greatly simplifies when one uses the identities:
\begin{align}
\tau^{\alpha_1\beta_1\gamma\delta\gamma'\delta'}(p)\tau^{\mu_1\nu_1\rho'\lambda'\rho\lambda}(q)L_{\alpha_1\beta_1\mu_1\nu_1}&=\frac{s^2}{2}L^{\gamma\delta\gamma'\delta'}L^{'\lambda'\rho\lambda}
\nonumber \\
L^{\gamma\delta\gamma'\delta'}L_{\gamma'\delta'\sigma'\eta'}L^{\sigma'\eta'\sigma\eta}&=4L^{\gamma\delta\sigma\eta}
\nonumber \\
L^{\gamma\delta\sigma\eta}\epsilon_{\gamma\delta}(p)\epsilon^*_{\sigma\eta}(p)&=2.
\label{identities}
\end{align}
That is, each ``rung" of the ladder gives a factor $\frac{s^2}{2}$, each propagator on the two ``legs" of the ladder gives a factor of 4, and there is an overall factor of 4 coming from the contractions of the polarization tensors. So, with our normalizations, the contraction of all the space-time indices in a ladder or crossed ladder diagram gives a factor of $\frac{1}{4}(8s^2)^n$, where $n$ is the number of gravitons exchanged.

Thus we have that:
\begin{align}
\mathcal{M}_2&=(-\frac{i\kappa}{2})^4s^4
\int \frac{d^4 k_1 d^4k_2}{(2\pi)^4}\delta(k_1+k_2+\Delta)\frac{1}{k_1^2+i\epsilon}
\nonumber \\
&\times \frac{1}{k_2^2+i\epsilon}\frac{1}{-2p\cdot k_1 +i\epsilon}
[\frac{1}{2q\cdot k_1 +i\epsilon}+\frac{1}{2q\cdot k_2 +i\epsilon}].
\end{align}
\subsubsection{$n-1$ Loop}
Let us now consider the case where $n$ gravitons are exchanged within the diagram. In this case after applying the identities in \eqref{identities} we have:
\begin{align}
\mathcal{M}_n&=-(-\frac{i\kappa}{2})^{2n}(-is^2)^n\int \frac{\prod_i^n d^4k_i}{(2\pi)^{4n-4}}\delta(\sum_i^n k_i+\Delta)
\nonumber \\
&\times \prod_i^n \frac{1}{k_i^2+i\epsilon}\prod_i^{n-1}\frac{1}{-2p\cdot k_i+i\epsilon}
 \sum_{\text{perms}}[\prod_i^{n-1} \frac{1}{2q\cdot k_i+i\epsilon}]
 \label{ngrav}
\end{align}
where the sum over permutations in the last line generate all of the distinct ladder and crossed ladder diagrams at this loop order.
\subsubsection{Exponentiation}
\label{sec:exponentiation}
We will now proceed to show that the expression for the $n-1$ loop scattering amplitude \eqref{ngrav} implies that the eikonal amplitude exponentiates. We will find it useful to evaluate this integral in light-cone coordinates. We then have:
\begin{align}
\mathcal{M}_n&=2s(-\frac{i\kappa}{2})^{2n}(\frac{is}{2})^n\int \frac{\prod_i^n dk_{i+}dk_{i-}d^2k_{i\perp}}{(2\pi)^{4n-4}}
 \delta(\sum_i^n k_{i+})\delta(\sum_i^n k_{i-})\delta(\sum_i^n k_{i\perp}+\Delta_{\perp})
\nonumber \\
&\times\prod_i^n\frac{1}{k_i^2+i\epsilon}\prod_i^{n-1}\frac{1}{k_{i-}-i\epsilon}
 \sum_{\text{perms}}[\prod_i^{n-1} \frac{1}{k_{i+}+i\epsilon}].
\end{align}
We will find it very useful to use the identity (A proof is outlined in Appendix \ref{secA} for completeness):
\begin{align}
&\delta(\omega_1+...+\omega_n)
\sum_{\text{Perms of }\omega_i}\frac{1}{\omega_{1}+i\epsilon}...\frac{1}{\omega_{1}+...+\omega_{n-1}+i\epsilon}
=(-2\pi i)^{n-1}\delta(\omega_1)...\delta(\omega_n)
\label{iden}
\end{align}
over the $k_{i+}$ coordinates to arrive at:
\begin{align}
\mathcal{M}_n&=2is(-\frac{i\kappa}{2})^{2n}(-\frac{s}{2})^n\int \frac{\prod_i^ndk_{i-}d^2k_{i\perp}}{(2\pi)^{3n-3}}
\delta(\sum_i^n k_{i-})\delta(\sum_i^n k_{i\perp}+\Delta_{\perp})
\prod_i^n\frac{1}{k_{i\perp}^2}\prod_i^{n-1}\frac{1}{k_{i-}-i\epsilon}.
\end{align}
We can then symmetrize the integrand in the $k_{i-}$ coordinates and again use \eqref{iden} to arrive at:
\begin{align}
\mathcal{M}_n&=-\frac{2s}{n!}(-\frac{i\kappa}{2})^{2n}(\frac{is}{2})^n\int \frac{\prod_i^nd^2k_{i\perp}}{(2\pi)^{2n-2}}
\delta(\sum_i^n k_{i\perp}+\Delta_{\perp})
\prod_i^n\frac{1}{k_{i\perp}^2}.
\end{align}
If we then Fourier transform the amplitude into impact parameter space we arrive at:
\begin{align}
\mathcal{N}_n&=\frac{1}{(2\pi)^2}\int d^{2}\Delta_\perp\mathcal{M}_ne^{-ib_\perp\cdot \Delta_\perp}
\nonumber \\
&=-\frac{2s}{n!}[\frac{-i\kappa^2 s}{8}\int \frac{d^2k_\perp}{(2\pi)^2}\frac{1}{k_{\perp}^2}e^{ib_\perp\cdot k_\perp}]^n.
\end{align}
So we see that when we sum over all $n$ we have for our full Fourier transformed amplitude:
\begin{align}
\mathcal{N}=-2s[e^{i\chi}-1]
\end{align}
where:
\begin{align}
\chi=-\frac{\kappa^2 s}{8}\int \frac{d^2k_\perp}{(2\pi)^2}\frac{1}{k_{\perp}^2}e^{ib_\perp\cdot k_\perp}.
\end{align}
This reproduces the well known result  that the amplitude exponentiates in impact parameter space.
\subsection{QCD}
\label{sec:qcd}

\begin{figure}
\begin{center}
\includegraphics[width=4in]{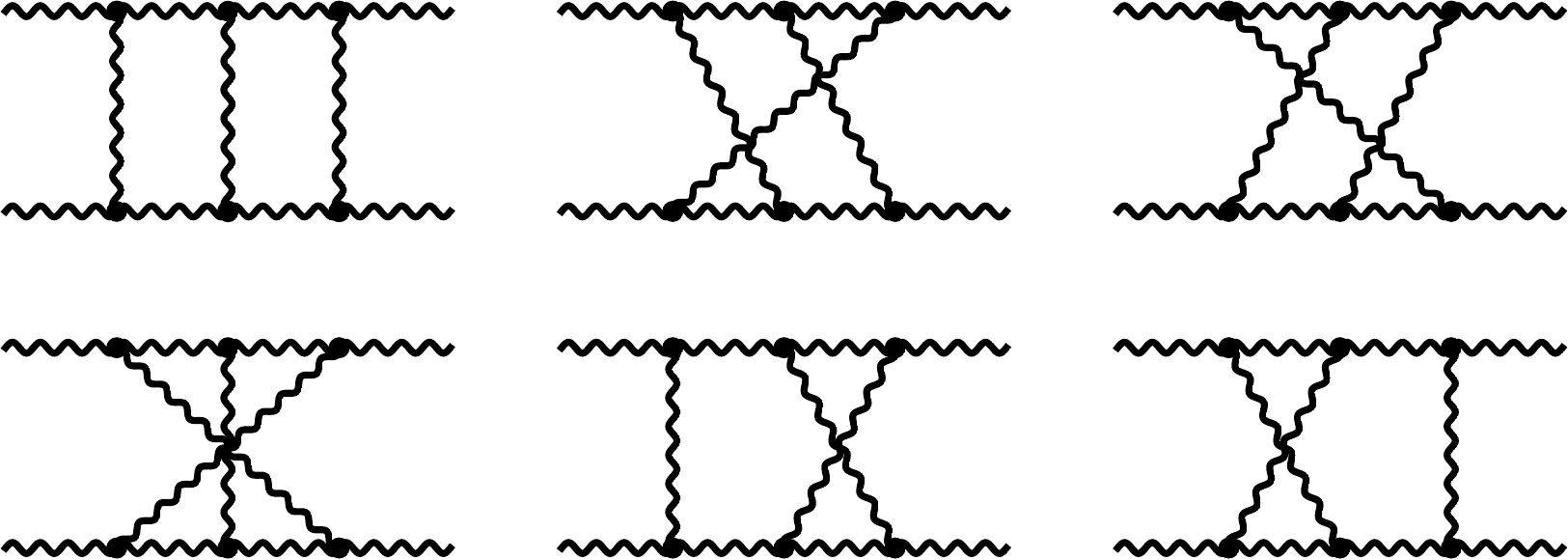}
\caption{The set of two-loop ladder diagrams relevant for the double copy relation in the eikonal approximation. For the same color factors the collinear divergences between these cancel.}
\label{fig6}
\end{center}
\end{figure}

In this section we identify the set of  gauge theory amplitudes at each loop order that can participate in a double copy relation with the gravity amplitudes derived in the previous section in the eikonal approximation. To be specific, we seek a double copy relation which relates the gravity amplitudes to two gauge theory amplitudes, both of which are in the eikonal approximation. As mentioned previously, an important constraint is that on the gauge theory side, all the collinear divergences have to cancel. This constraint becomes relevant because for gauge theories there are more diagrams in the leading eikonal approximation than those discussed in the previous section for the gravity case.

While it is well known that only ladder and crossed ladder diagrams contribute to the graviton scattering amplitude at leading eikonal order, this is not the case for nonabelian gauge theories such as QCD. The eikonal approximation in QCD has been extensively studied in the context of the Reggeization phenomenon. Ref. \cite{ChengWu} depicts all the diagrams to the third loop order and summarizes the result of investigations in this area. For an interesting attempt to investigate high-energy QCD scattering using less standard methods, see ref. \cite{Verlinde}, who model the scattering process as a simplified two dimensional effective field theory.

One of the big differences between QCD and gravity at the one loop order stems from the fact that the corresponding box and crossed box diagrams (figures \ref{fig2} and \ref{fig3}) come with different color factors. Each of the corresponding Feynman integrals contributes a $\ln s$ term coming from a  collinear divergence and for gravity these $\ln s$ terms cancel between the two diagrams. However, this cancellation is prevented in QCD because of the different color factors for the two diagrams.

For example, at the one loop level the box graph has a color factor of 
$(T^aT^b)_{ij}(T^aT^b)_{kl}$, whereas the the crossed box gives $(T^aT^b)_{ij}(T^bT^a)_{kl}$. Here, $T^a$ are the matrices in the adjoint representation and $(ij)$ and $(kl)$ denote the color indices of the external gluons. We may rewrite the color factor of the crossed box as $(T^aT^b)_{ij}(T^bT^a)_{kl}=(T^aT^b)_{ij}(T^aT^b)_{kl} + \frac{3}{2}(T^a)_{ij}(T^a)_{kl}$. Thus the sum of the two diagrams now has a piece proportional to the same color factor $(T^aT^b)_{ij}(T^aT^b)_{kl}$ and a piece which has the same color factor as the tree level diagram. The piece proportional to $(T^aT^b)_{ij}(T^aT^b)_{kl}$ is just like the contribution in gravity for which the $\ln s$ terms cancel, and the term proportional to $\frac{3}{2}(T^a)_{ij}(T^a)_{kl}$ contains a $\ln s$ term which together with similar pieces from higher orders containing higher powers of $\ln s$ terms then Reggeizes the one gluon exchange amplitude \cite{ChengWu,Lam}. The key point now is that it is only the first type of contribution which is relevant for the double copy relation. This is because the leading numerator factors in the eikonal approximation have no dependence on loop momenta, so these logarithms will persist even once the double copy conjecture is applied. However, as collinear singularities are known to be absent in gravitational amplitudes, we will need to systematically restrict our attention to a subset of the QCD amplitudes that has no such logarithms. When the double copy relation is applied to the full gauge theory the cancellation of those $\ln s$ terms which are responsible for gluon Reggeization  proceeds through terms other than those given by the leading eikonal approximation and the exact mechanism for this is beyond the scope of this paper. (The fact that all the collinear divergences do cancel in the gauge theory side once the exact double copy relations are applied has been explicitly checked at fixed order for some theories in \cite{Dixon}). As we discuss later in this section, this argument can be easily extended to higher loop orders and the net result is the following: In order to eliminate the $\ln s$ contributions which arise from collinear divergences, we restrict our attention only to those diagrams in the scattering amplitude that are same ones as considered for the case of gravitational eikonal scattering and have the common color structure of an uncrossed ladder diagram (for instance, the color factors associated with figure \ref{fig2} and the first diagram of figure \ref{fig6} at orders $g^4$ and $g^6$, respectively). The fact that the $\ln{s}$ dependence in such terms cancel through sixth-order in the coupling constant was seen in \cite{ChengWu,Lam}. We show this cancellation endures to all orders in the following sections.

\begin{figure}
\begin{center}
\includegraphics[width=2.5in]{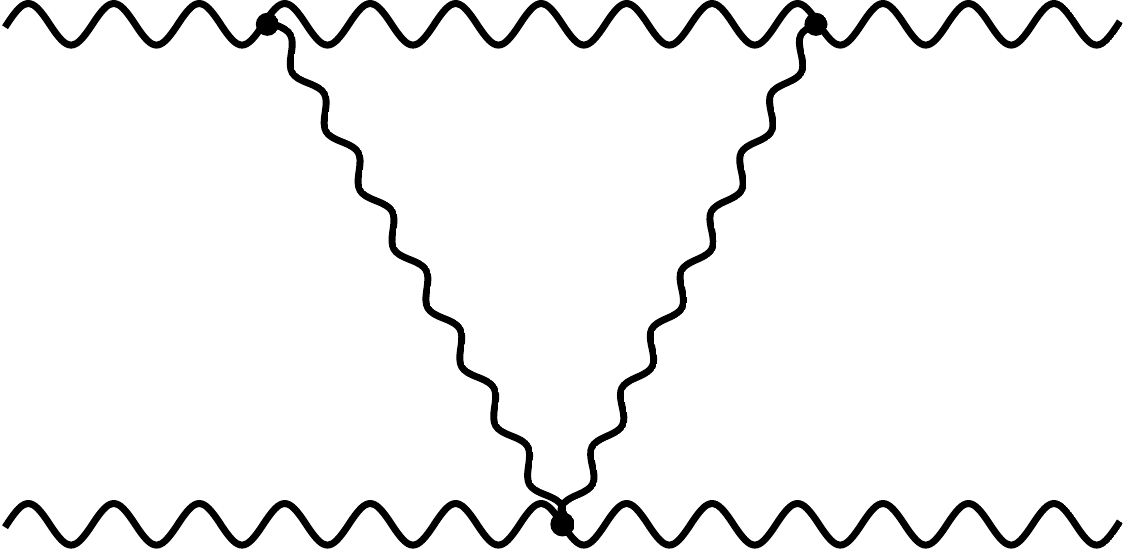}
\caption{The seagull diagram.}
\label{fig5}
\end{center}
\end{figure}

\begin{figure}
\begin{center}
\includegraphics[width=2.5in]{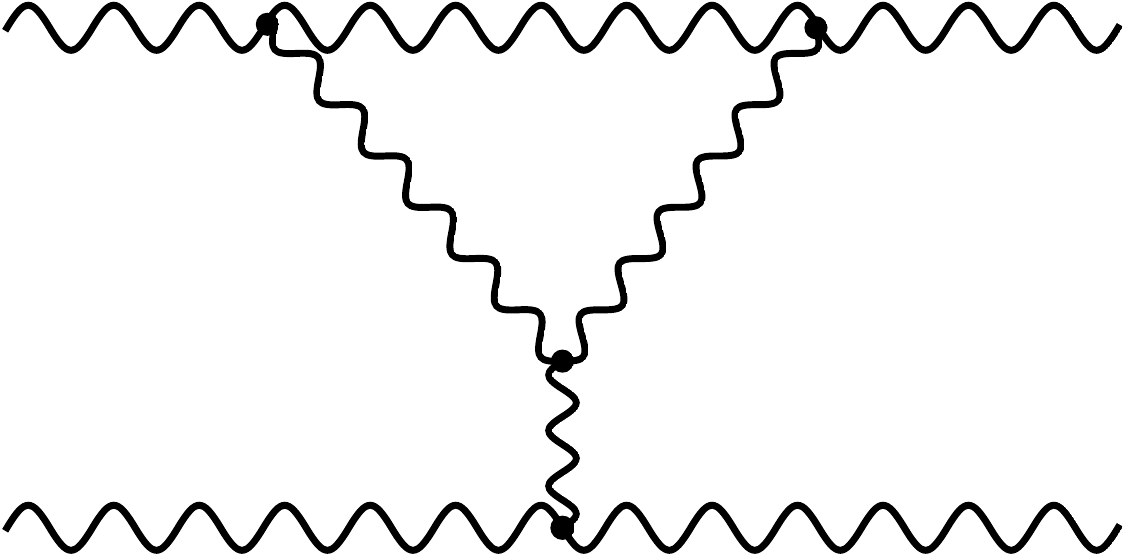}
\caption{The triangle diagram.}
\label{fig4}
\end{center}
\end{figure}

Before we proceed to a systematic discussion of the class of diagrams of interest to us for checking the double copy relation in the eikonal approximation, we would like to comment on the contributions of the seagull and the triangle type diagrams shown in figure \ref{fig5} and figure \ref{fig4} respectively.

Let us begin with diagrams with seagull type interactions (for example figure \ref{fig5}). We will show that these are subleading in the eikonal approximation even for the case of gauge theories. In order to see this, note that the Feynman integral for this diagram is:
\begin{align}
I_{\text{sea}}&=\int \frac{dk_{1+}dk_{2+} dk_{1-}dk_{2-} d^2k_{1\perp}d^2k_{2\perp}}{(2\pi)^4}
\delta(k_{1+}+k_{2+})\delta(k_{1-}+k_{2-})
\nonumber \\
&\times\delta^2(k_{1\perp}+k_{2\perp}+\Delta_{\perp})\frac{1}{k_{1+}k_{2+}q_-}\frac{1}{k_{1-}-\frac{k_{1\perp}^2}{k_{1+}}+\frac{i\epsilon}{k_{1+}}}
\frac{1}{k_{2-}-\frac{k_{2\perp}^2}{k_{2+}}+\frac{i\epsilon}{k_{2+}}}\frac{1}{k_{1+}+i\epsilon}
\end{align}
We have assumed there are two vertices attaching onto the bottom leg but switching to the case where two vertices attach to the upper leg is trivial. Using the delta function over $k_{2-}$ we have:
\begin{align}
I_{\text{sea}}&=-\int \frac{dk_{1+}dk_{2+} dk_{1-} d^2k_{1\perp}d^2k_{2\perp}}{(2\pi)^4}\delta(k_{1+}+k_{2+})
\nonumber \\
&\times\delta^2(k_{1\perp}+k_{2\perp}+\Delta_{\perp})\frac{1}{k_{1+}k_{2+}q_-}\frac{1}{k_{1-}-\frac{k_{1\perp}^2}{k_{1+}}+\frac{i\epsilon}{k_{1+}}}
\frac{1}{k_{1-}+\frac{k_{2\perp}^2}{k_{2+}}-\frac{i\epsilon}{k_{2+}}}\frac{1}{k_{1+}+i\epsilon}
\end{align}
Now let us consider the integration over $k_{1-}$. Due to the delta function $\delta(k_{1+}+k_{2+})$ we see that both poles in $k_{1-}$ must lie on the same side of the real axis. Now, it is well known (for a pedagogical discussion see \cite{Sterman}) that infrared sensitive effects only arise from pinch singular points. Because of this, it is easy to see that the above does not contain any infrared effects. Thus a Feynman integral with eikonalized propagators will not have any leading-order contributions for diagrams with a seagull interaction.

Note that we can extend the above reasoning for the case of diagrams with a three point interaction at one loop order as in figure \ref{fig4}. This is because the only difference between the Feynman integral of the seagull diagram (figure \ref{fig5}) and the triangle diagram (figure \ref{fig4}) are numerator factors of $k_{i+}$, which do not alter the presented argument.

Keeping these things in mind, in this section we give the expressions for the matrix elements of diagrams contributing to high-energy gluon-gluon scattering at leading eikonal order. As discussed earlier, we only retain terms with the color structure corresponding to uncrossed ladder diagrams and keep only the lowest order coefficient of each color factor. We explicitly show what we mean by this below.
\subsubsection{Tree Level}

We will start by stating the tree level result. This is straightforward at leading eikonal order since any of the polarization vectors contracted with its own momenta gives zero, which kills all of the factors in the three-gluon vertex at leading eikonal order except for one from each vertex.:
\begin{align}
\mathcal{A}_{1}&=i \frac{g^2}{\Delta^2}g^{\alpha\beta}g^{\rho \sigma}(p+p')\cdot (q+q')f_{acb}f_{ade}
\epsilon_\alpha(p)\epsilon_\rho(q)\epsilon^*_\beta(p')\epsilon^*_\sigma(q')
\nonumber \\
&\approx i \frac{2g^2s}{\Delta^2}F_{1}
\label{QCDTree}
\end{align}
where we have written the color factor as $F_1 = (T^a)_{ij}(T^a)_{kl}$.
\subsubsection{One-Loop}

Let us now move on to the one-loop level. Much like in the graviton scattering case, at leading eikonal order the kinematic factors in the vertices have a very simple structure.

This is due to the fact that the eikonalized three-gluon vertex gives a factor of $2iT^ag_{\alpha\beta}p_\mu$ on the top leg and a factor of $2iT^ag_{\alpha\beta}q_\mu$ on the bottom leg. This is because all other vertex factors are proportional to momenta that will contract with either itself or their corresponding polarization vector. As the factors of $g^{\alpha\beta}$ in the vertices contract to give a factor of $\epsilon \cdot \epsilon^*=1$, we are left with a factor of $-2s$ for each ``rung" of the ladder, multiplied by a diagram dependent color factor. Thus at one-loop we have for the matrix element:

\begin{align}
\mathcal{A}_{2}&=4s^2g^4\int \frac{d^4 k_1 d^4k_2}{(2\pi)^4}\delta(k_1+k_2+\Delta)\frac{1}{k_1^2+i\epsilon}
\nonumber \\
&\times \frac{1}{k_2^2+i\epsilon}\frac{1}{-2p\cdot k_1 +i\epsilon}
[\frac{F_2}{2q\cdot k_1 +i\epsilon}+\frac{F_2'}{2q\cdot k_2 +i\epsilon}]
 \label{QCDOneLoop}
\end{align}
where the sum in the second line represents the box and crossed box diagrams and $F_2=(T^aT^b)_{ij}(T^aT^b)_{kl}$ and $F_2'=(T^aT^b)_{ij}(T^bT^a)_{kl}$ are their respective color factors. Recall that we can use the group theory relation:
\begin{align}
F_2'=F_2+\frac{3}{2} F_1
\end{align}
To arrive at:
\begin{align}
\mathcal{A}_{2}&=4s^2g^4F_2\int \frac{d^4 k_1 d^4k_2}{(2\pi)^4}\delta(k_1+k_2+\Delta)\frac{1}{k_1^2+i\epsilon}
\nonumber \\
&\times \frac{1}{k_2^2+i\epsilon}\frac{1}{-2p\cdot k_1 +i\epsilon}
[\frac{1}{2q\cdot k_1 +i\epsilon}+\frac{1}{2q\cdot k_2 +i\epsilon}]
+\mathcal{O}(g^4)F_1
\label{oneloopeqn}
\end{align}
where we have ignored the term proportional to $F_1$ since it has an uncanceled collinear divergence. Note that a convenient bookkeeping device is to keep only the term that is lowest order in the coupling constant for each relevant color factor. In this case, the contribution to $F_1$ from \eqref{oneloopeqn} is higher order than the contribution from \eqref{QCDTree}.

At the two-loop level we will consider only the set of diagrams in figure \ref{fig6}. Our prescription is to then commute the color factors so that in the end we only retain a contribution with a color factor of the first diagram in figure \ref{fig6}.
\subsubsection{$n-1$ Loop}

We now state the expression for the matrix element at the $n-1$ loop level. Knowing that we get a factor of $-2s$ times a color factor for each ``rung" of the ladder makes this task very straightforward. We have:
\begin{align}
\mathcal{A}_n&=-g^{2n}(-2is)^n\int \frac{\prod_i^n d^4k_i}{(2\pi)^{4n-4}}\delta(\sum_i^n k_i+\Delta)
\nonumber \\
&\times \prod_i^n \frac{1}{k_i^2+i\epsilon}\prod_i^{n-1}\frac{1}{-2p\cdot k_i+i\epsilon}
 \sum_{\text{perms}}[\prod_i^{n-1} \frac{c_{\text{perm}}}{2q\cdot k_i+i\epsilon}]
\end{align}
where the sum over permutations that generate all the distinct ladder and crossed ladder diagrams at this order now include the distinct color factor for each of these diagrams, which we write as $c_{\text{perm}}$. Note that as we did in the 1-loop case, we can commute the terms in each distinct color factor so that they all are reduced to $F_n=(T^{a_1}T^{a_2}....T^{a_n})_{ij}(T^{a_1}T^{a_2}....T^{a_n})_{kl}$, the color factor corresponding to the $n-1$ loop uncrossed ladder diagram, plus terms that have color factors corresponding to lower loop diagrams but are ignored due to having higher power dependence on the coupling constant. These latter terms have collinear divergences and are therefore of no interest to us.  Thus we can write:
\begin{align}
\mathcal{A}_n&=-g^{2n}F_{n}(-2is)^n\int \frac{\prod_i^n d^4k_i}{(2\pi)^{4n-4}}\delta(\sum_i^n k_i+\Delta)
\nonumber \\
&\times \prod_i^n \frac{1}{k_i^2+i\epsilon}\prod_i^{n-1}\frac{1}{-2p\cdot k_i+i\epsilon}
 \sum_{\text{perms}}[\prod_i^{n-1} \frac{1}{2q\cdot k_i+i\epsilon}]
+\mathcal{O}(g^{2n})F_{i\le n-1}
\end{align}
Note that it is clear that the above expression will not have any collinear divergences since we can proceed as we did in section \ref{sec:exponentiation} and integrate over the $+$ and $-$ components of all the loop momenta, leaving only transverse components.

\subsection{Double Copy Relation}
\label{sec:double}

We have seen in the last section that by retaining only the lowest order coefficient of color structures corresponding to uncrossed ladder diagrams, we can eliminate all collinear logarithms in the QCD eikonal scattering amplitude. Now that we have identified the correct subset of the amplitude to consider, a double copy relation is seen immediately, as:

\begin{align}
\frac{(-i)^{n-1}}{g^{2n}}\mathcal{A}_n&=-i\tilde{F}_{n}s^n\int \frac{\prod_i^n d^4k_i}{(2\pi)^{4n-4}}\delta(\sum_i^n k_i+\Delta)
\prod_i^n \frac{1}{k_i^2+i\epsilon}
\nonumber \\
&\times\prod_i^{n-1}\frac{1}{-2p\cdot k_i+i\epsilon}\sum_{\text{perms}}[\prod_i^{n-1} \frac{1}{2q\cdot k_i+i\epsilon}]
\end{align}
where we have rescaled $F_n\rightarrow \frac{\tilde{F}_n}{(-2)^n}$ as prescribed by \cite{BCJ2}. We then see from \eqref{ngrav} that:
\begin{align}
\frac{(-i)^n}{(\kappa/2)^{2n}}\mathcal{M}_n&=-s^{2n}\int \frac{\prod_i^n d^4k_i}{(2\pi)^{4n-4}}\delta(\sum_i^n k_i+\Delta)
 \prod_i^n \frac{1}{k_i^2+i\epsilon}
\nonumber \\
&\times\prod_i^{n-1}\frac{1}{-2p\cdot k_i+i\epsilon}\sum_{\text{perms}}[\prod_i^{n-1} \frac{1}{2q\cdot k_i+i\epsilon}]
\end{align}
so we see that if we replace the color factor $\tilde{F}_{n}$ with the numerator factor $-is^n$ the gravitational result is recovered.

\section{Connection to Shockwaves}
\begin{figure}
\begin{center}
\includegraphics[width=2.5in]{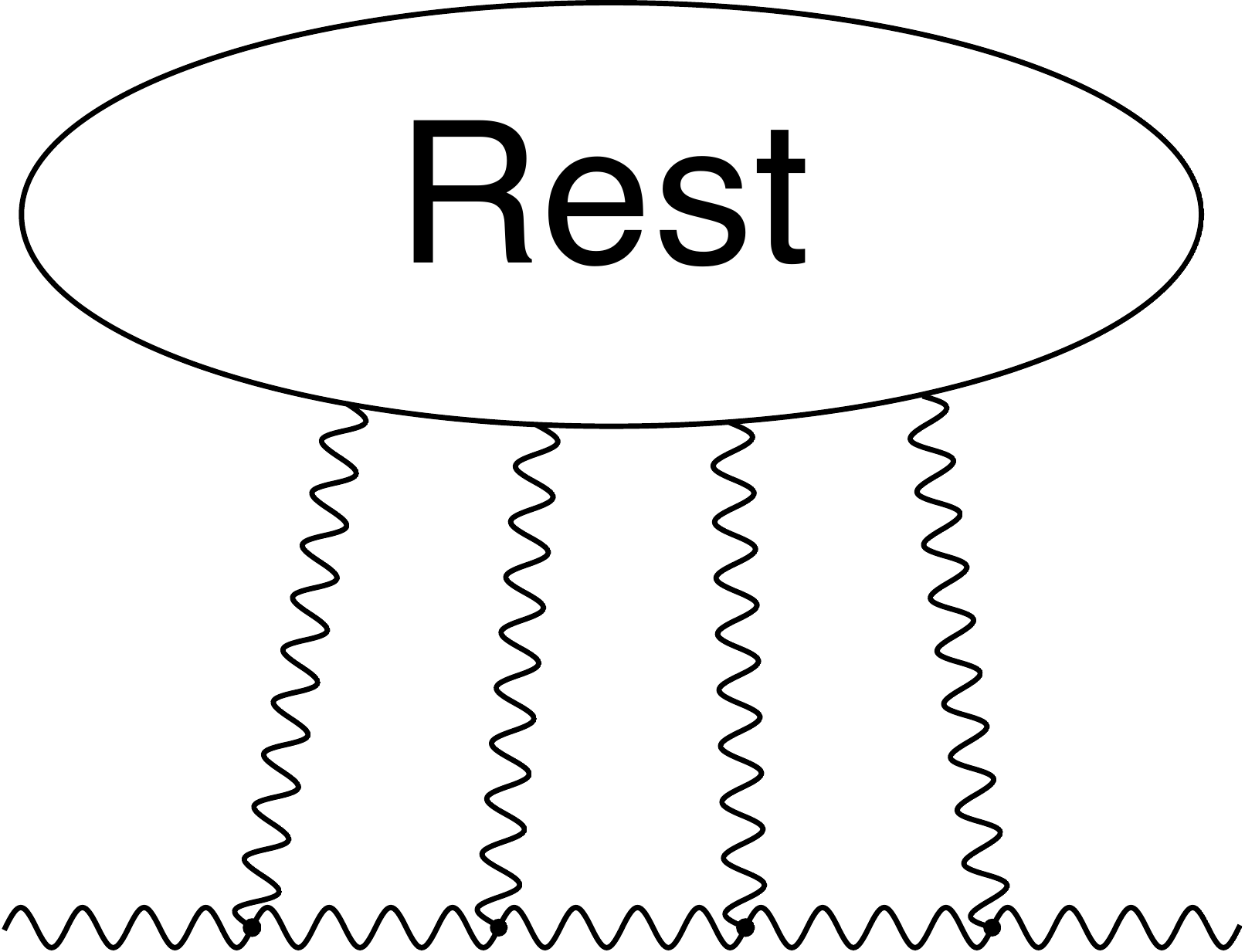}
\caption{A high energy graviton/gluon interacting with the rest of the diagram through eikonalized graviton/gluon exchange. We show the case with $n=4$ gravitons/gluons for purposes of illustration.}
\label{Rest1}
\end{center}
\end{figure}

\begin{figure}
\begin{center}
\includegraphics[width=2.5in]{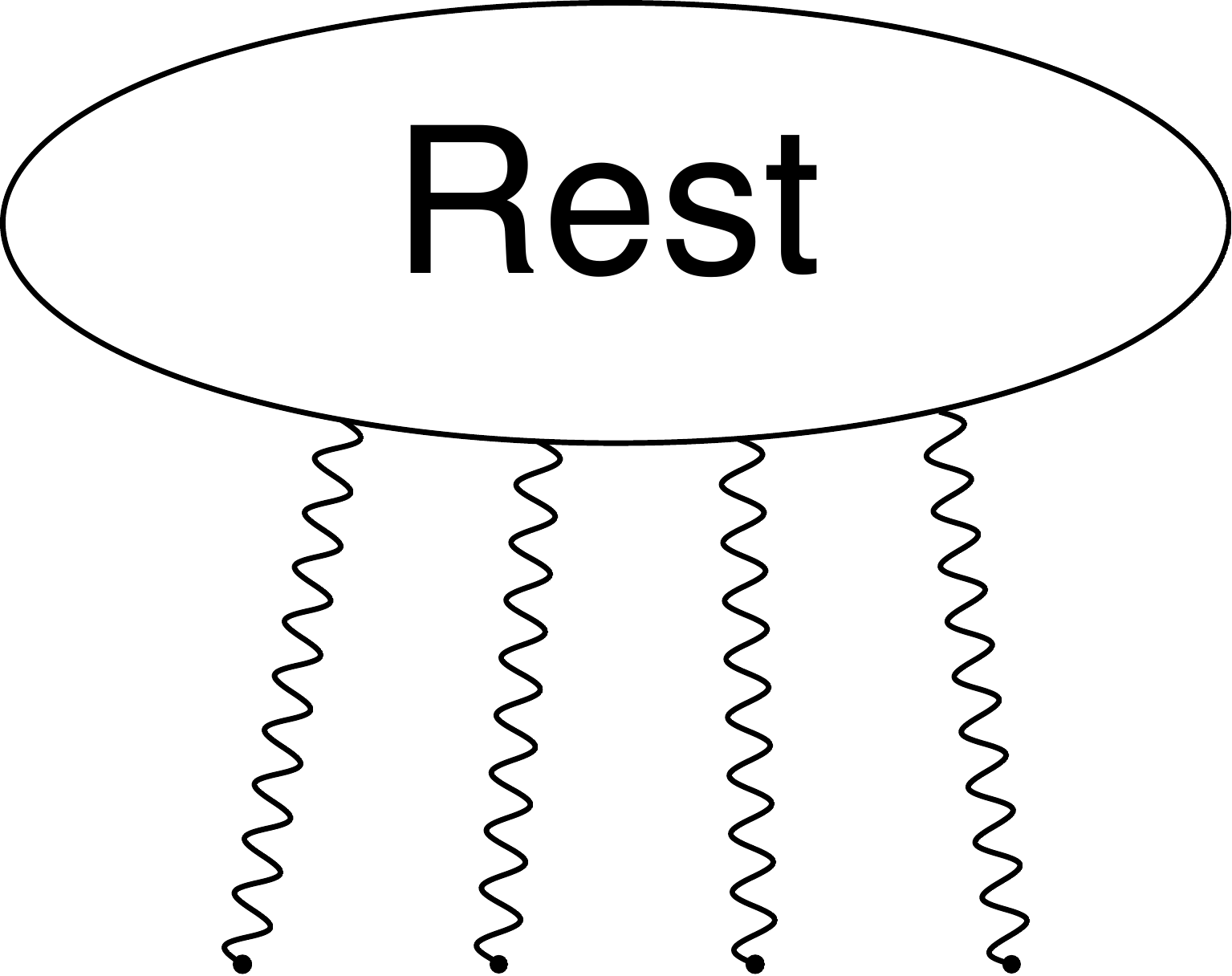}
\caption{Interaction of the rest of the diagram with an external tensor/vector field. We show the case with $n=4$ gravitons/gluons for purposes of illustration.}
\label{Rest2}
\end{center}
\end{figure}

In this section we will explore the possibility of replacing a fast moving particle in interaction with the rest of an arbitrary Feynman diagram by an external potential. In particular, 
we will present a field theoretic calculation that directly illustrates the connection between the eikonal approximation and shockwaves. This is done by comparing the expressions for the interaction of an arbitrary diagram through eikonalized graviton or gluon exchange with a relativistic graviton/gluon line (figure \ref{Rest1}) to that of its interaction with external tensor/vector field sources (figure \ref{Rest2}).

\subsection{Aichelberg-Sexl Metric}
\label{sec:gravshock}

We first start with the gravitational case, for which the shockwave is described by the Aichelberg-Sexl metric. Let us consider an arbitrary diagram $R$ interacting with a relativistic graviton line by exchanging $n$ eikonalized gravitons. In this case using similar methods to those used in section \ref{sec:grav} we see that the matrix element corresponding to this process is:
\begin{align}
\mathcal{M}_n^{(A)}&=-i(i\kappa)^{n}\int \frac{\prod_i^n d^4k_i}{(2\pi)^{4n-4}}\delta^4(\sum_i^n k_i+\Delta)\prod_i^nq_{\mu_i}q_{\nu_i}
\nonumber \\
&\times \prod_i^n \frac{1}{k_i^2+i\epsilon}
 \sum_{\text{perms}}[\prod_i^{n-1} \frac{1}{2q\cdot k_i+i\epsilon}]
 R^{\{\mu_i,\nu_i\}}(\{p_i\}, \{k_i\})
\end{align}
where $R^{\{\mu_i,\nu_i\}}(\{p_i\}, \{k_i\})$ denotes the expressions for the rest of the diagram. The $p_i$ are the momenta going into and out of rest, and we are still defining $\Delta=p'-p=q-q'$, where $p$ is the net momentum going into $R$ and $p'$ is the net momentum going out. We can now simplify our result using \eqref{iden}:
\begin{align}
\mathcal{M}_n^{(A)}&=\kappa^{n}2q_0\int \frac{\prod_i^n d^4k_i}{(2\pi)^{3n-3}}\delta^3(\sum_i^n {\bf k_i+\Delta})\prod_i^nq_{\mu_i}q_{\nu_i}
\nonumber \\
&\times \prod_i^n \frac{\delta(2q\cdot k_i)}{k_i^2+i\epsilon}
 R^{\{\mu_i,\nu_i\}}(\{p_i\}, \{k_i\})
\end{align}
Let us now compare this result to the case where this same arbitrary diagram is interacting $n$ times with an external tensor field:
\begin{align}
h_{\mu\nu}=\int\frac{d^4k}{(2\pi)^4}e^{i(k\cdot x)}C_{\mu\nu}(k)
\label{field}
\end{align}
The net result is:
\begin{align}
\mathcal{M}_n^{(B)}&=\int\frac{\prod_i^n d^4k_i}{(2\pi)^{4n-4}}\delta^4(\sum_i^n k_i+\Delta)
\nonumber \\
&\times C_{\mu_1\nu_1}(k_1)...C_{\mu_n\nu_n}(k_n)R^{\{\mu_i,\nu_i\}}(\{p_i\}, \{k_i\}),
\end{align}
Let us choose:
\begin{align}
C_{\mu\nu}(k)=2\kappa\pi q_\mu q_\nu \delta(2q\cdot k)\frac{1}{k^2+i\epsilon}
\label{coeff}
\end{align}
Note that:
\begin{align}
&\delta^4(p'-p+k_1+...+k_n)\delta(2q\cdot k_1)...\delta(2q\cdot k_n)
\nonumber \\
&=\delta(p'^0-p^0+k^0_1+...+k^0_n)\delta^3({\bf p'-p+k_1+...+k_n})
\delta(2q\cdot k_1)...\delta(2q\cdot k_n)
\nonumber \\
&=\delta(p'^0-p^0+{\bf v_q\cdot(p-p')})\delta^3({\bf q-q'+k_1+...+k_n})
\delta(2q\cdot k_1)...\delta(2q\cdot k_n)
\nonumber \\
&=2q_0\delta(2q\cdot(p-p'))\delta^3({\bf q-q'+k_1+...+k_n})
\delta(2q\cdot k_1)...\delta(2q\cdot k_n)
\end{align}
where ${\bf v_q}\equiv \frac{{\bf q}}{q_0}$ and we have used the fact that there are delta functions that impose the condition $k_i^0={\bf v_q\cdot k_i}$ and that $\bf{p-p'=q'-q=\sum k_i}$. Now, $\delta(p'^0-p^0+k^0_1+...+k^0_n)=2q_0\delta(2q\cdot(p-p'))$ can be pulled out of the ${\bf{k}}$ integrations and we get:
\begin{align}
\mathcal{M}_n^{(B)}&=2\pi \delta(2q\cdot(p-p'))\mathcal{M}_n^{(A)}
\end{align}
so we see that the effect of the high energy line is the same as that of the external field \eqref{field} with $C_{\mu\nu}(k)$ given by \eqref{coeff}. Note that this field can be written as:
\begin{align}
h_{\mu\nu}&=\kappa q_\mu q_\nu\int\frac{d^4k}{(2\pi)^3}e^{i(k\cdot x)}\delta(2q\cdot k)\frac{1}{k^2+i\epsilon}
\nonumber \\
&=\frac{\kappa}{2q_0}q_\mu q_\nu\int\frac{d^3{\bf k}}{(2\pi)^3}e^{-i{\bf k}\cdot ({\bf r-v_q}t)}\frac{1}{\bf (v_q\cdot k)^2-k^2}
\end{align}
If we take ${\bf q}$ in the z direction, we have:
\begin{align}
h_{\mu\nu}&=-\frac{\kappa}{2} q_0u_\mu u_\nu\int\frac{d^2{\bf k_\perp} dk_3}{(2\pi)^3}e^{-i{\bf k_\perp}\cdot {\bf r_\perp}}
e^{-ik_3(z-v_qt)}\frac{1}{{\bf k_\perp^2}+k_3^2(1-v_q^2)}
\end{align}
Let us take the limit as $v_{q}\rightarrow 1$ ($q_0$ is held fixed), in which case the external field becomes:
\begin{align}
h_{\mu\nu}=-\frac{\kappa}{2} q_0u_\mu u_\nu\delta(z-t)\int\frac{d^2{\bf k_\perp}}{(2\pi)^2}e^{-i{\bf k_\perp}\cdot {\bf r_\perp}}\frac{1}{{\bf k_\perp^2}}
\label{h}
\end{align}
where $u$ is the 4-velocity corresponding to $q$. The equation of motion for the de Donder gauge we are using is:
\begin{align}
\Box(h_{\alpha\beta}-\frac{1}{2}\eta_{\alpha\beta}h)=-\frac{\kappa}{2}T_{\alpha\beta}
\end{align}
Note in the limit $v_q\rightarrow1$, $h=0$. We then have using $h_{\alpha\beta}$ from \eqref{h}:
\begin{align}
\Box h_{\alpha\beta}&=(\partial_{t+z}\partial_{t-z}-\partial_{r_\perp}^2)h_{\alpha\beta}
\nonumber \\
&=-\frac{\kappa}{2} q_0u_\alpha u_\beta\delta(z-t)\delta(y)\delta(x)
\end{align}
Thus the energy momentum tensor corresponding to our external field is:
\begin{align}
T_{\alpha\beta}=q_0u_\alpha u_\beta\delta(z-t)\delta(y)\delta(x)
\end{align}
Which is precisely the energy momentum tensor of a massless particle, which will give the Aichelburg-Sexl metric once the Einstein equation is solved for.

We can also directly apply a gauge transformation to \eqref{h} to show that it is indeed the Aichelberg-Sexl metric. The key fact to note here is that:
\begin{align}
\partial_x \int\frac{d^2{\bf k_\perp}}{(2\pi)^2}e^{-i{\bf k_\perp}\cdot {\bf r_\perp}}\frac{1}{{\bf k_\perp^2}}=\partial_x \frac{-1}{4\pi}\ln(r_\perp^2)=-\frac{1}{2\pi}\frac{x}{x^2+y^2}
\end{align}
Because of this, when we make the gauge transformation:
\begin{align}
h'_{\mu\nu}= h_{\mu\nu}+\partial_{\mu}\eta_\nu+\partial_\nu\eta_\mu
\end{align}
with:
\begin{align}
\eta_-=\frac{\kappa}{4} q_0\Theta(z-t)[\int\frac{d^2{\bf k_\perp}}{(2\pi)^2}e^{-i{\bf k_\perp}\cdot {\bf r_\perp}}\frac{1}{{\bf k_\perp^2}}+\frac{1}{4\pi}\ln(r_\perp^2)]
\end{align}
and all other components of $\eta_\mu$ being zero, we have $\partial_x \eta=\partial_y\eta=0$. Thus, by choosing this gauge parameter we ensure that we only change the $h_{--}$ component with this gauge transformation, which is the only non-zero component.

After this gauge transformation we arrive at:
\begin{align}
\kappa h'_{\mu\nu}=4G q_0u_\mu u_\nu\delta(z-t)\ln(r_\perp^2)
\end{align}
which is precisely the standard expression for the Aichelberg-Sexl metric (for example, see eqn. 1 of \cite{DT}).
\subsection{Gauge Shockwave}
\label{sec:gaugeshock}

Here we discuss how one can arrive at a gauge shockwave using methods analogous to those used to arrive at the Aichelberg-Sexl metric.

Let us first consider an arbitrary diagram exchanging $n$ gluons with a high energy gluon line. Using similar methods to section \ref{sec:qcd} we find that the matrix element corresponding to this process is:
\begin{align}
\mathcal{A}_n^{(A)}&=-i(2g)^{n}\int \frac{\prod_i^n d^4k_i}{(2\pi)^{4n-4}}\delta^4(\sum_i^n k_i+\Delta)
\nonumber \\
&\times \prod_i^nq_{\mu_i}\prod_i^n \frac{1}{k_i^2+i\epsilon}
 \sum_{\text{perms}}[\prod_j^{n-1} \frac{d_n^{\{a_i\}}}{2q\cdot k_j+i\epsilon}]
R^{\{\mu_i\}}_{\{a_i\}}(\{p_i\}, \{k_i\})
\end{align}
Note that each permutation that generates a new diagram will have a different color factor $d_n^{\{a_i\}}$ which will be contracted with the rest of the diagram. However, we can again commute the color matrices so that they all reduce to a common color factor plus a suppressed term that has a color factor of a lower order diagram and would contain potential collinear divergences. Again, we ignore terms of this latter type to arrive at:

\begin{align}
\mathcal{A}_n^{(A)}&=-i(2g)^{n}D_n^{\{a_i\}}\int \frac{\prod_i^n d^4k_i}{(2\pi)^{4n-4}}\delta^4(\sum_i^n k_i+\Delta)
\nonumber \\
&\times \prod_i^nq_{\mu_i}\prod_i^n \frac{1}{k_i^2+i\epsilon}
 \sum_{\text{perms}}[\prod_i^{n-1} \frac{1}{2q\cdot k_i+i\epsilon}]
R^{\{\mu_i\}}_{\{a_i\}}(\{p_i\}, \{k_i\})
\end{align}
where $D_n^{\{a_i\}}$ is the color factor that corresponds to the diagrams where $n$ gluons are exchanged with the rest of the diagram in an uncrossed fashion. Typically it will be of the form,
$D_n^{\{a_i\}}=(T^{a_1}T^{a_2}....T^{a_n})_{ij}$. Note that there is a similar chain of the product of the $T^a$ included inside $R^{\{\mu_i\}}_{\{a_i\}}(\{p_i\}, \{k_i\})$.

Then, after using \eqref{iden} we have:
\begin{align}
\mathcal{A}_n^{(A)}&=(-2ig)^{n}2q_0D_n^{\{a_i\}}\int \frac{\prod_i^n d^4k_i}{(2\pi)^{3n-3}}\delta^3(\sum_i^n {\bf k_i+\Delta})
\nonumber \\
&\times \prod_i^nq_{\mu_i}\prod_i^n \frac{\delta(2q\cdot k_i)}{k_i^2+i\epsilon}
R^{\{\mu_i\}}_{\{a_i\}}(\{p_i\}, \{k_i\})
\end{align}
Let us now compare this expression to that obtained by letting this arbitrary diagram interact $n$ times with an external vector field:
\begin{align}
A_\mu^{(a)}=\int\frac{d^4k}{(2\pi)^4}e^{i(k\cdot x)}B_{\mu}^{(a)}(k)
\end{align}
In this case, we find that the resulting amplitude is:
\begin{align}
\mathcal{A}_n^{(B)}&=\int\frac{\prod_i^n d^4k_i}{(2\pi)^{4n-4}}\delta^4(\sum_i^n k_i+\Delta)
B_{\mu_1}^{(a_1)}(k_1)...B_{\mu_n}^{(a_n)}(k_n)R^{\{\mu_i\}}_{\{a_i\}}(\{p_i\}, \{k_i\})
\end{align}
Note that if we make the identification:
\begin{align}
B_\mu^{(a)}(k)=-4ig\pi q_\mu D_1^{(a)}\delta(2q\cdot k)\frac{1}{k^2+i\epsilon}
\end{align}
we see that:
\begin{align}
\mathcal{A}_n^{(B)}&=(-2ig)^n4\pi q_0\delta(2q\cdot(p-p'))
\int\frac{\prod_i^n d^4k_i}{(2\pi)^{3n-3}}\delta^3(\sum_i^n {\bf k_i+\Delta})
\nonumber \\
&\times \prod_i^n D_1^{(a_i)}\prod_i^nq_{\mu_i}\prod_i^n \frac{\delta(2q\cdot k_i)}{k_i^2+i\epsilon}
R^{\{\mu_i\}}_{\{a_i\}}(\{p_i\}, \{k_i\})
\end{align}
Note that as $\prod_i^n D_1^{(a_i)}=D_n^{\{a_i\}}$, we have:

\begin{align}
\mathcal{A}_n^{(B)}&=2\pi \delta(2q\cdot(p-p'))\mathcal{A}_n^{(A)}
\end{align}
as desired.
Going through the exact same procedure as the gravitational case, we then find that:
\begin{align}
A_\mu^{(a)}(k)=-ig u_\mu D_1^{(a)} \delta(z-t)\int\frac{d^2{\bf k_\perp}}{(2\pi)^2}e^{-i{\bf k_\perp}\cdot {\bf r_\perp}}\frac{1}{{\bf k_\perp^2}}
\label{A}
\end{align}
If we insert this vector field into the Yang-Mills equation:
\begin{align}
D^{\mu}F_{\mu\nu}^{(a)}=j_\nu^{(a)}
\end{align}
we find that the corresponding current is:
\begin{align}
j_{\mu}^{(a)}=-ig u_\mu D_1^{(a)}\delta(z-t)\delta(x)\delta(y)
\end{align}
Note this is exactly the source that gives rise to a QED shockwave,  with the identification of $igD_1^{(a)}$ being the electromagnetic charge. 

As we did for the gravitational case we can also instead show the equivalence using a gauge transformation:
\begin{align}
A'_\mu=A_\mu+\partial_\mu \Omega
\end{align}
with:
\begin{align}
\Omega=igD_1^{(a)}\Theta(t-z)\int\frac{d^2{\bf k_\perp}}{(2\pi)^2}e^{-i{\bf k_\perp}\cdot {\bf r_\perp}}\frac{1}{{\bf k_\perp^2}}
\end{align}
we have:
\begin{align}
&A_0=A_z=0;
\nonumber \\
&A_{\perp}=-\frac{igD_1^{(a)}}{4\pi}\Theta(t-z)\nabla \ln(r_\perp^2)
\end{align}
which is exactly the QED result with the identification of $igD_1^{(a)}$ being the electromagnetic charge. So we have seen that deriving a shockwave in the eikonal limit causes it to take an abelian form, even for the case of nonabelian gauge theories such as QCD.
\subsection{Relationship Between The Two Shockwaves}
\label{sec:doubleshock}

It is interesting to note that the double copy relation similar to that which we showed in section \ref{sec:double} can also be seen in the two shockwaves \eqref{h} and \eqref{A}. For the full scattering amplitude we looked at the gravitational quantity $\frac{(-i)^n}{(\kappa/2)^{2n}}\mathcal{M}_n$. The analogous quantity to look at for the gravitational shockwave is:
\begin{align}
\frac{1}{\kappa}h_{\mu\nu}=-q_\mu q_\nu \frac{\delta(z-t)}{2q_0}\int\frac{d^2{\bf k_\perp}}{(2\pi)^2}e^{-i{\bf k_\perp}\cdot {\bf r_\perp}}\frac{1}{{\bf k_\perp^2}}
\end{align}
since we have accounted for factors corresponding to graviton couplings and numerator factors of the propagators that reside in ``rest". Note that we have identified $\frac{\delta(z-t)}{2q_0}$ as being a propagator as this piece was derived solely from non-numerator contributions.

On the gauge theory side, the quantity we looked at for the full scattering amplitude was $\frac{(-i)^{n-1}}{g^{2n}}\mathcal{A}_n$. The analogous quantity to look at for the shockwave is:
\begin{align}
\frac{1}{g}A_\mu=iq_\mu \tilde{D}_1^{(a)}\frac{\delta(z-t)}{2q_0}\int\frac{d^2{\bf k_\perp}}{(2\pi)^2}e^{-i{\bf k_\perp}\cdot {\bf r_\perp}}\frac{1}{{\bf k_\perp^2}}
\end{align}
where we have accounted for a factor of $g$ that corresponds to gluon couplings that reside in ``rest". We have also rescaled the color factor to $\tilde{D}_1^{(a)}=-\frac{1}{2}D_1^{(a)}$ (instead of scaling by $-\frac{i}{\sqrt{2}}$) since there are two factors of $D_1^{(a)}$ that need to be accounted for in the $n=1$ case that the single external gauge field corresponds to (one factor in the gauge field and another factor residing in the attachment onto ``rest").

We then easily see that if we make the replacement $\tilde{D}_1^{(a)}\rightarrow i q_\mu$ in the gauge shockwave we clearly recover the gravitational result. In concluding this section we would like to point out that this double copy relation is only obvious in our choice of gauge and is greatly obscured in any other gauge.
\section{Discussion}

We have shown that there exists a double copy relation between eikonalized gravity and gauge theory amplitudes, as long as we consider only the lowest order contribution to the color factors corresponding to completely uncrossed ladder diagrams. This restriction is necessary as the inclusion of other contributions to the amplitude will result in collinearly divergences which are known to cancel in gravitational amplitudes. This is a feature particular to the eikonal approximation, as in this limit the numerator factors have no dependence on the loop momenta which does not allow one to change the integrals found in gauge theories once the double copy conjecture is applied.

An interesting consequence of this double copy relation between the eikonalized amplitudes is that the corresponding shockwave solutions have a double copy relation as well. While both the gauge and gravity shockwaves had been calculated previously using completely classical methods, our method directly shows the relationship between eikonal amplitudes and shockwaves. Furthermore, the double copy relation between the two shockwaves had not previously been seen. This is because this double copy relation is only apparent in a particular choice of gauge, which was naturally selected by our method of calculation.

One aspect of this analysis we would like to comment on is that we did not need to consider gravity coupled to a dilaton or an anti-symmetric tensor in order to arrive at a double copy relation in this kinematic regime at leading eikonal order. Such a coupling is necessary for a double copy relation in the unrestricted kinematic region as discussed in \cite{BCJ1, BCJ2}. However, it is to be expected that we would not need to consider these couplings as in the soft regime scalars decouple due to power counting arguments. This reasoning is also why ghosts do not need to be considered in the soft limit \cite{DeWitt}. Arguments for the decoupling of scalars and anti-symmetric tensors also apply at the tree level, as is well known.

\begin{acknowledgments} 
We would like to thank the SLAC theory group and in particular, Lance Dixon, for hospitality during August 2012 where part of this work was done. We also thank Lance Dixon for discussions. 
R.S. and R.A. are supported by a grant from the U.S. Department of Energy, grant number DE-SC 0007859.
\end{acknowledgments}
\appendix
\section{Proof of an Identity Used in The Paper}
\label{secA}
In this appendix, for completeness, we present a proof of the identity \eqref{iden} used in the main text. This identity is a very useful one for applications of the eikonal approximation \cite{ChengWu}.
\begin{align}
&\delta(\omega_1+...+\omega_n)
\sum_{\text{Perms of }\omega_i}\frac{1}{\omega_{1}+i\epsilon}...\frac{1}{\omega_{1}+...+\omega_{n-1}+i\epsilon}
\nonumber \\
&=(-2\pi i)^{n-1}\delta(\omega_1)...\delta(\omega_n)
\label{iden3}
\end{align}
We will use the representations:
\begin{align}
&\delta(\omega_1+\omega_2+...+\omega_n)=\frac{1}{2\pi}\int dt_n \prod_{j=1}^n e^{-i\omega_j t_n}
\nonumber \\
& \frac{1}{\omega+i\epsilon}=-i\int d\tau \theta(\tau) e^{i(\omega+i\epsilon)\tau}
\end{align}
To rewrite the left hand side of \eqref{iden3} as:
\begin{align}
&\delta(\omega_1+...+\omega_n)
\sum_{\text{Perms of }\omega_i}\frac{1}{\omega_{1}+i\epsilon}...\frac{1}{\omega_{1}+...+\omega_{n-1}+i\epsilon}
\nonumber \\
&=(-i)^{n-1}\frac{1}{2\pi}\sum_{\text{Perms over } i}\int dt_n 
[\prod_{j=1}^{n-1}d\tau_j\theta(\tau_j)\exp(i\sum_{k=1}^j(\omega_{i_k}+i\epsilon)\tau_j)][\prod_{j=1}^n e^{-i\omega_j t_n}]
\end{align}
Now let us make the coordinate transformation $\tau_i=t_{i+1}-t_i$. The Jacobian of this transformation is 1, so we find:
\begin{align}
&\delta(\omega_1+...+\omega_n)
\sum_{\text{Perms of }\omega_i}\frac{1}{\omega_{1}+i\epsilon}...\frac{1}{\omega_{1}+...+\omega_{n-1}+i\epsilon}
\nonumber \\
&=(-i)^{n-1}\frac{1}{2\pi}\sum_{\text{Perms of }\omega_i}\int dt_n...dt_1
\theta(t_n-t_{n-1})...\theta(t_2-t_1)
\nonumber \\
&\times e^{i\omega_{i_1}(\tau_1+\tau_2+...+\tau_{n-1}-t_n)}e^{i\omega_{i_2}(\tau_2+\tau_3+...+\tau_{n-1}-t_n)}...e^{i\omega_{i_{n-1}}(\tau_{n-1}-t_n)}e^{i\omega_{i_{n}}(-t_n)}
\nonumber \\
&=(-i)^{n-1}\frac{1}{2\pi}\sum_{\text{Perms of }\omega_i}\int dt_n...dt_1
\theta(t_n-t_{n-1})...\theta(t_2-t_1)\exp(-i\sum_{j=1}^n\omega_{i_j}t_j)
\nonumber \\
&=(-i)^{n-1}\frac{1}{2\pi}\int dt_n...dt_1\exp(-i\sum_{j=1}^n\omega_{i_j}t_j)
\nonumber \\
&=(-2\pi i)^{n-1}\delta(\omega_1)...\delta(\omega_n)
\end{align}
So we see that \eqref{iden3} does indeed hold true.


\begin{thebibliography}{99}

\bibitem{BCJ1}
Z.~Bern, J. J. M.~Carrasco and H.~Johansson, {\it New relations for gauge-theory amplitudes}, 
  {\it Phys. Rev.} {\bf D 78} (2008) 085011 [arXiv:0805.3993 [hep-ph]].
  
\bibitem{BCJ2}
Z.~Bern, J. J. M.~Carrasco and H.~Johansson, {\it Perturbative Quantum Gravity as a Double Copy of Gauge Theory},
{\it Phys. Rev. Lett.} {\bf 105} (2010) 061602 [arXiv:1004.0476 [hep-th]].

\bibitem{thooft}
  G.  't Hooft, {\it Graviton dominance in ultra-high-energy scattering}, {\it Phys. Lett.} {\bf B} {\bf{198}} (1987) 61.
  
\bibitem{Veneziano1}
D.~Amati, M.~ Ciafaloni and G.~ Veneziano, {\it Superstring collisions at planckian energies}, {{\it Phys. Lett.} {\bf{B 197}} (1987) 81}.
  
\bibitem{Veneziano2}
D.~Amati, M.~ Ciafaloni and G.~ Veneziano, {\it Classical and Quantum Gravity Effects from Planckian Energy Superstring Collisions}, {{\it Int. J. Mod. Phys.} {\bf{A 3}},
  (1988) 1615}.
  
\bibitem{Veneziano3}
D.~Amati, M.~ Ciafaloni and G.~ Veneziano, {\it Higher-order gravitational deflection and soft bremsstrahlung in planckian energy superstring collisions}, {{\it Nucl. Phys.} {\bf{B 347}} (1990) 550}.
  
\bibitem{Veneziano4}
D.~Amati, M.~ Ciafaloni and G.~ Veneziano, {\it Planckian scattering beyond the semiclassical approximation}, {{\it Phys. Lett.} {\bf{B 289}} (1992) 87}.
  
\bibitem{Veneziano5}
D.~Amati, M.~ Ciafaloni and G.~ Veneziano, {\it Effective action and all-order gravitational eikonal at planckian energies}, {{\it Nucl. Phys.} {\bf{B 403}} (1993) 707}.
  
  \bibitem{GGM}
S.~Giddings, D.~Gross and A.~Maharana, {\it Gravitational effects in ultrahigh-energy string scattering}, {{\it 
  Phys. Rev.} {\bf D 77} (2008) 046001 [arXiv:0705.1816 [hep-th]]}.

\bibitem{Giddings}
S.~Giddings, {\it The gravitational S-matrix: Erice lectures}, (2011)[arXiv:1105.2036 [hep-th]].

\bibitem{Weinberg}
S.~Weinberg, {\it Infrared Photons and Gravitons}, {{\it
  Phys. Rev.} {\bf 140} (1965) B516}.

\bibitem{Us}
R.~Akhoury, R.~Saotome and G.~Sterman, {\it Collinear and soft divergences in perturbative quantum gravity}, {{\it
  Phys. Rev.} {\bf D 84} (2011) 104040} [arXiv:1109.0270 [hep-th]].

\bibitem{Beneke}
M.~Beneke and G. Kirilin, {\it Soft-collinear gravity}, {{\it JHEP} {\bf 1209} (2012) 066} [arXiv:1207.4926 [hep-ph]].

\bibitem{White}
S.~Oxburgh and C. D.~White, {\it BCJ duality and the double copy in the soft limit}, (2012) [arXiv:1210.1110 [hep-th]].

\bibitem{AS}
P. C.~Aichelburg and R. U.~Sexl, {\it On the gravitational field of a massless particle}, {{\it
Gen. Rel. Grav.} {\bf 2}, (1970) 303}.

\bibitem{DT}
T.~Dray and G.~'t Hooft, {\it The gravitational shock wave of a massless particle}, {{\it Nucl. Phys.} {\bf B 253} (1985) 173}

\bibitem{Jackiw}
R.~Jackiw, D.~Kabat and M.~Ortiz, {\it Electromagnetic fields of a massless particle and the eikonal}, {{\it Phys. Lett.} {\bf B 277} (1992) 148} [hep-th/9112020]

\bibitem{ChengWu}
H.~Cheng and T. T.~Wu, {\it Expanding Protons: Scattering at High Energies}, M.I.T. Press, Cambridge, MA (1987)

\bibitem{Verlinde}
H.~Verlinde and E.~Verlinde, {\it QCD at High Energies and Two-Dimensional Field Theory}, (1993) [hep-th/9302104]

\bibitem{Lam}
Y. J.~Feng, O.~Hamidi-Ravari and C. S.~Lam, {\it Cut diagrams for high energy scatterings}, {{\it Phys. Rev.} {\bf D 54} (1996) 3114} [hep-ph/9604429]

\bibitem{Dixon}
C.~Boucher-Veronneau and L.~J.~Dixon, {\it N $\ge$ 4 Supergravity Amplitudes from Gauge Theory at Two Loops}, {\it JHEP} {\bf 1112} (2011) 046
 [arXiv:1110.1132 [hep-th]].

\bibitem{Sterman}
  G.~F.~Sterman,
{\it An Introduction to quantum field theory},
University Press, Cambridge, UK (1993).

\bibitem{DeWitt}
B. S.~DeWitt, {\it Quantum Theory of Gravity. III. Applications of the Covariant Theory}, {{\it Phys. Rev.} {\bf 160} (1967) 1113}.


\end{thebibliography}
\end{document}